\documentclass[aip,twocolumn,superscriptaddress,10pt,reprint]{revtex4-1}
\usepackage{graphicx}
\usepackage{multirow}
\usepackage{amsmath}

\begin{document}

\title{Evidence for anisotropic dielectric properties of monoclinic hafnia using\\ high-resolution TEM valence electron energy-loss spectroscopy and\\ \textbf{\textit{ab initio}} time-dependent density-functional theory}
\author{C. Guedj}
\affiliation{CEA, LETI, MINATEC Campus, 17 rue des Martyrs, 38054 Grenoble, France.}
\author{L. Hung}
\affiliation{LSI, CNRS, CEA and \'Ecole Polytechnique, Palaiseau, France.}
\affiliation{European Theoretical Spectroscopy Facility (ETSF).}
\author{A. Zobelli}
\affiliation{LPS, CNRS and Univ. Paris sud, Orsay, France.}
\author{P. Blaise}
\affiliation{CEA, LETI, MINATEC Campus, 17 rue des Martyrs, 38054 Grenoble, France.}
\affiliation{European Theoretical Spectroscopy Facility (ETSF).}
\author{F. Sottile}
\affiliation{LSI, CNRS, CEA and \'Ecole Polytechnique, Palaiseau, France.}
\affiliation{European Theoretical Spectroscopy Facility (ETSF).}
\author{V. Olevano}
\affiliation{Institut NEEL, CNRS and Univ. Grenoble Alpes, Grenoble, France.}
\affiliation{European Theoretical Spectroscopy Facility (ETSF).}

\date{\today}

\begin{abstract}

The effect of nanocrystal orientation on the energy loss spectra of monoclinic hafnia (m-HfO$_2$) is measured by high resolution transmission electron microscopy (HRTEM) and valence energy loss spectroscopy (VEELS) on high quality samples.
For the same momentum-transfer directions, the dielectric properties are
also calculated \textit{ab initio} by time-dependent density-functional theory (TDDFT).
Experiments and simulations evidence anisotropy in the dielectric properties of m-HfO$_2$, most notably with the direction-dependent oscillator strength of the main bulk plasmon.
The anisotropic nature of m-HfO$_2$ may contribute to the differences among VEELS spectra reported in literature.
The good agreement between the complex dielectric permittivity extracted from VEELS with nanometer spatial resolution, TDDFT modeling, and past literature demonstrates that the present HRTEM-VEELS device-oriented methodology is a possible solution to the difficult nanocharacterization challenges given in the International Technology Roadmap for Semiconductors.
\end{abstract}


\maketitle

With the downscaling of microelectronic and optoelectronic devices, accurate metrology at the nanoscale has become an important objective for the microelectronic industry.
At the same time, the International Technology Roadmap for Semiconductors categorizes the ``measurement of complex material stacks and interface properties, including physical and electrical properties'' as a ``difficult challenge'' for $\sim$16 nm CMOS technology nodes\cite{itrs}.
The characterization of high-$\kappa$ gate stacks (mostly based on hafnia-based dielectrics) is particularly complicated due to the length scales at which electronic properties are determined.
These new challenges for characterization and metrology arise not only from the introduction of thinner and more complex materials and stacks, but also from the need to discern physical properties at an increasing spatial resolution. To develop nanocharacterization protocols that are independent of materials stacks and integration design, even more advanced methods are required. 
To our knowledge, (valence) electron energy-loss spectroscopy (V)EELS is the only technique capable of measuring dielectric and optical properties\cite{Stoger-Pollach} (complex refractive index), and chemical properties\cite{WangWang} (composition, atomic bonding) at the same time and with nanometer spatial resolution, when all effects are properly taken into account \cite{retardation, retardation2}.

In this paper, we use the energy filtered TEM-VEELS technique (also known as EFTEM SI),\cite{VerbeeckVanDyck} in a high-resolution transmission electron microscope (HRTEM) to simultaneously obtain the structural and spectroscopic properties of $P2_1/c$ m-HfO$_2$ with nanometric spatial resolution. 
HfO$_2$ is a prominent high-$\kappa$ material used in various applications like MIM capacitors\cite{MIM}, resistive memories (OxRRAM)\cite{OxRRAM} or optical coatings \cite{Coatings}.
To this purpose, the dielectric properties of m-HfO$_2$ corresponding to the different crystal configuration and orientations that can be grown in an electronic device must be precisely measured, a task for which HRTEM-VEELS is particularly suited.
After detailed nanostructural modeling of HRTEM measurements using quantitative image simulations, we obtaine VEELS spectra for various well identified m-HfO$_2$ crystal orientations.
For the same momentum-transfer directions, we also calculate \textit{ab initio} time-dependent density-functional theory (TDDFT)\cite{TDDFT} energy-loss spectra.
For the calculated TDDFT spectra, we use the random-phase approximation (RPA) and include local-field effects\cite{Adler,Wiser} on top. 
This level of theory allows us to reproduce, interpret, and even predict experimental energy-loss spectra\cite{WeisskerSerrano}. 
With its reliability and predictivity, TDDFT is a valuable complement to experimental applied research on high-$\kappa$ materials for electronic devices.

For the dielectric properties and spectra of m-HfO$_2$, we find a significant dependence on the crystal direction.
The analysis of the main oscillators of the EELS spectra shows that the change in lattice orientation mostly affects the strength of the main bulk plasmon excitation at $\sim$16 eV.
A significant anisotropy of $\sim$10\% has been found also on the dielectric constant.

\begin{figure*}
\includegraphics[width=0.9\textwidth]{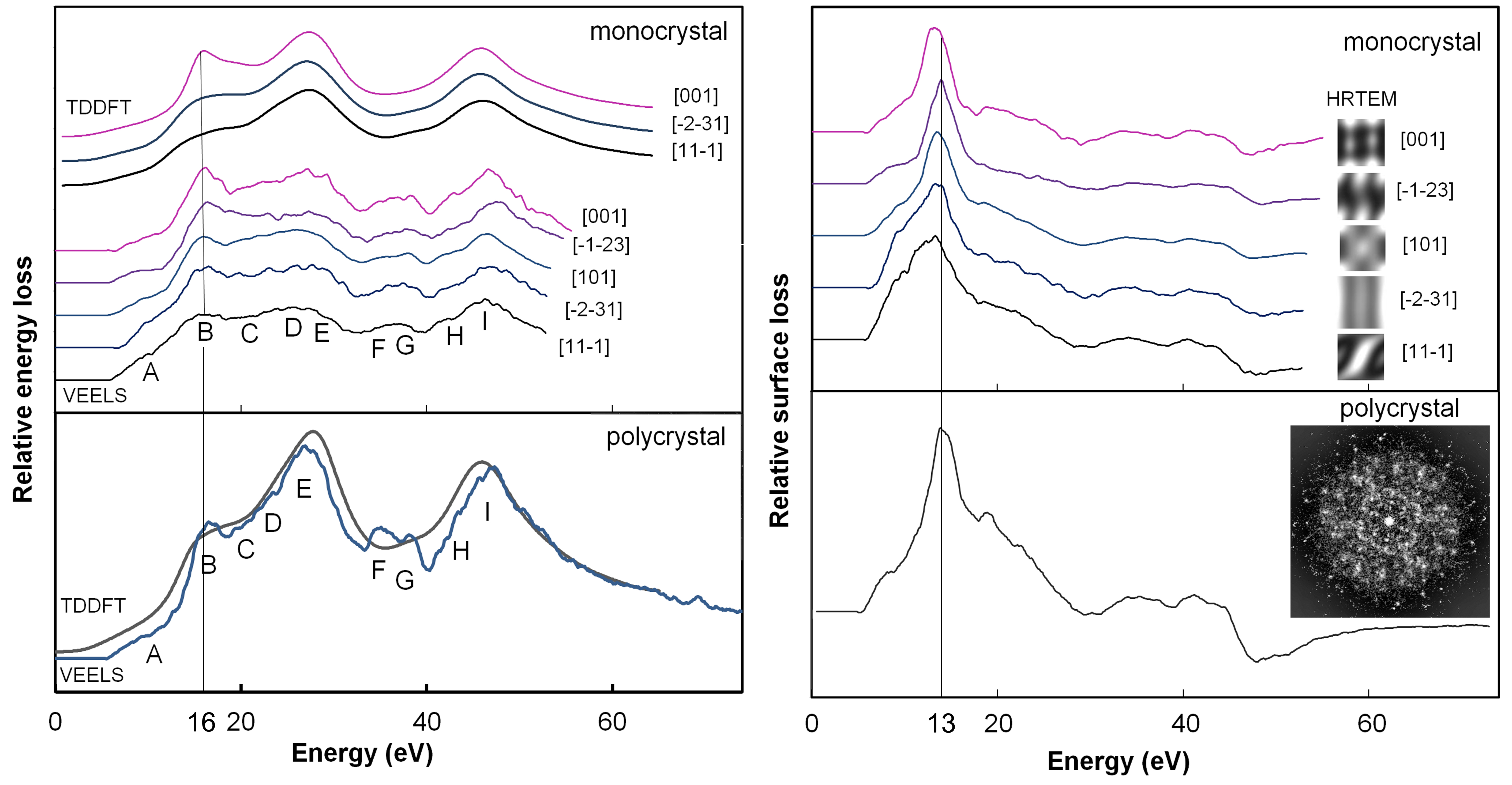}
\caption{
Valence bulk energy loss (left) and surface-loss spectra (right) deconvoluted from measured VEELS spectra of m-HfO$_2$  in the case of monocrystals (top) and polycrystals (bottom) at several momentum-transfer directions, in comparison with ab initio TDDFT simulations. 
The TEM images of monocrystals and the selected area diffraction pattern of the polycrystal are also provided.
Curves in top panels are vertically shifted to facilitate comparison.
}
\label{anisotropy-figure}
\end{figure*}

\paragraph{Experiment}
HfO$_2$ films are grown by atomic layer deposition (ALD) on 200 mm p-Si(100) wafers. 
Before deposition, substrates are treated with a diluted HF solution to remove any native oxides. 
ALD takes place in an ASM Pulsar 2000\textsuperscript{\texttrademark} module at $350\,^\circ\mathrm{C}$ using alternating pulses of HfCl$_4$ and H$_2$O, with N$_2$ as a carrier gas.
The introduction of H$_2$O vapor is used to desorb HCl at the growing surface, and the cycles are repeated sequentially to reach a thickness sufficient to maximize the crystal quality and to minimize the surface losses. A final annealing at $650\,^\circ\mathrm{C}$ is performed to crystallize HfO$_2$ and minimize the amount of oxygen vacancies.
A high Tauc-Lorentz\cite{JellisonModine} band gap ($\sim 5.9$) is measured by spectroscopic ellipsometry and VEELS, which is consistent\cite{HildebrandtKurian} with a low level of oxygen vacancies of this fully oxidized HfO$_2$ layer. 
This layer is therefore representative of a good quality dielectric material used in the microelectronic industry. Additional measurements are performed on high grade m-HfO$_2$ powders for verification.

Cross-sectional electron microscopy and diffraction experiments are performed in a JEOL 2010 FEF transmission electron microscope (TEM) operated at 200 kV at a magnification of 800 k, with an energy step of 0.1 eV between each image acquisition. The lowest achievable collection and convergence angles are used (few mrads) to minimize experimental momentum dispersion. The measured energy resolution is typically around 1.4 eV. For verification, complementary results are obtained with the Cs-corrected Titan microscope operated at 200 keV in STEM and TEM modes. About 80 millions spectra are acquired over 16 different samples to check the consistency of the results and to optimize the protocols of data acquisition and analysis. Samples are prepared with a Strata\textsuperscript{\texttrademark} 400 DualBeam\textsuperscript{\texttrademark} FIB/STEM system using Ga+ ions energies ranging from 30 keV down to 2 keV. An improvement in the quality of HRTEM-VEELS data is obtained by selective lift-off of superficial amorphous species by HF etching. Experimental data are corrected using the guidelines provided by Schaffer \textit{et al}.\cite{SchafferKothleitner} 
TEM lamella thickness is optimized ($< 40$ nm) to avoid the need for multiple scattering deconvolution processing, but not too thin ($> 15$ nm) to avoid excessive surface effects. Quantitative spectra are extremely difficult to obtain because of the numerous sources of variability due to instrumentation, sample preparation and data analysis. The zero-loss (elastic) contribution is removed from a reference VEELS spectrum acquired simultaneously in the vacuum region closest to the measured region of interest. The quality and reproducibility of the deconvolution process is verified by bandgap analysis of millions of spectra. The TEM approach is particularly convenient for absolute comparison of 2 neighbour grains with different orientations, because the data acquisition is simultaneous for both nanocrystals and the sources of instrumental variability can be deconvoluted more efficiently.
Fortunately, m-HfO$_2$ appears to be very stable under e-beam irradiation. The Kramers-Kronig analysis\cite{Egerton} is then performed on the single scattering distributions using classical routines available in the Digital Micrograph\textsuperscript{\texttrademark} environment to provide complex permittivities, energy-loss functions and surface-loss functions versus local nanostructure.   


\paragraph{Theory}
Numerical calculations\cite{conv-param} are carried out within the framework of density-functional theory (DFT) using a planewave and pseudopotential implementation in a two-step approach:
First, the ground state atomic structure and electronic density of m-HfO$_2$ is computed by static DFT\cite{DFT} using the local-density approximation (LDA)\cite{LDA} and the code ABINIT\cite{abinit}.
The calculated m-HfO$_2$ lattice parameters are in good agreement with our and literature\cite{WangLi} experimental values.
We use a Hf pseudopotential that includes semicore 4$f$, 5$s$ and 5$p$ electrons in valence since they contribute to excitations in the studied energy range. 
Second, the energy-loss and the dielectric function are calculated by linear-response TDDFT\cite{TDDFT} using the DP\cite{dp} code.
The inclusion of local-field effects has been found to be crucial to correctly reproduce the HfO$_2$ energy-loss function.
To compare with VEELS, TDDFT spectra are convoluted with a broadening of 1.5 eV, of the order of the experimental energy resolution.

\paragraph{Identification of crystal structures and orientations}

A careful analysis of the crystal structure is necessary since hafnia has several phases depending on pressure\cite{LegerAtouf} or growth method \cite{cubic,tetragonal,Hirabayashi,orthorhombic}. 
The monoclinic phase\cite{monoclinic} (space group P2$_1$/c) is the most stable in ambient conditions.

The simulated diffraction patterns of the different phases are often very similar, therefore the distinction between the HfO$_2$ polymorphs is difficult.
The comparison between experimental and simulated defocus-thickness series usually provides an identification of the phase and orientation of hafnia. In the worse cases, an unambiguous identification is provided by exit wave reconstruction techniques using the True Image FEI \textregistered Software\cite{TIFEI}.

\begin{figure}
\centerline{\includegraphics[width=\columnwidth]{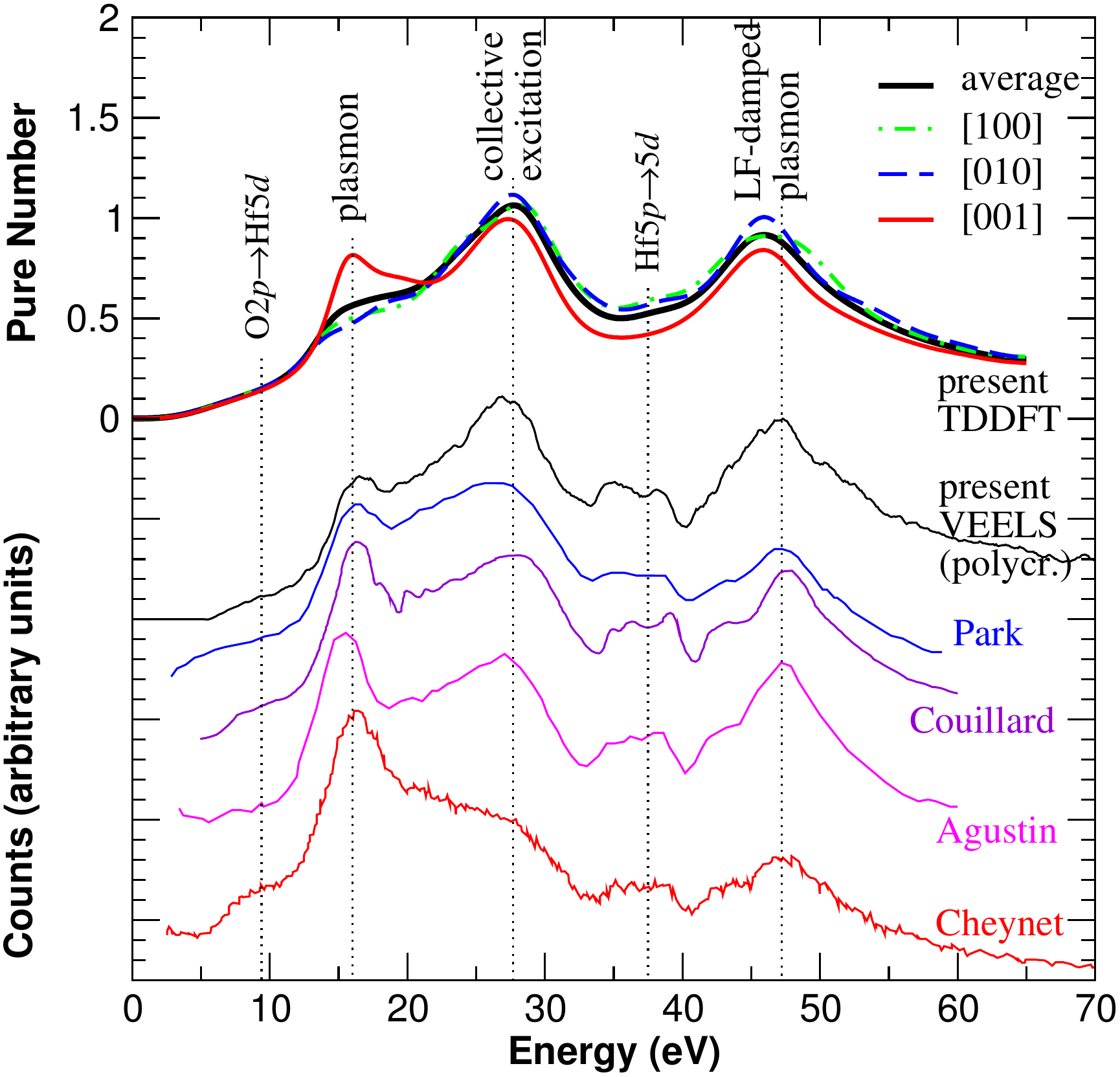}}
\caption{
Energy-loss spectra of monoclinic HfO$_2$.
From top to bottom: TDDFT RPA with local fields calculation as in the [100] (dot-dashed green), [010] (dashed blue) and [001] (thin solid red) directions and their average (thick solid black line);
VEELS spectra as measured by us (polycrystalline, black), by Park and Yang (blue)\cite{ParkYang}, Couillard et al. (violet)\cite{Couillard}, Agustin et al. (magenta)\cite{Agustin}, and Cheynet et al. (red)\cite{CheynetPokrant}.
Experimental curves are vertically shifted to facilitate comparison.
\label{figure-biblio}
}
\end{figure}

\paragraph{Results: energy-loss anisotropy}

VEELS spectra of single crystalline m-HfO$_2$ measured at negligible transferred momentum oriented along five different directions are presented in the top left of Fig.~\ref{anisotropy-figure}.
The spectra usually display ten apparent features labelled from A to I, and respectively located at around 10, 16, 20, 23, 27, 35, 38, 44 and 47 eV, although some peaks overlap and could be considered as a broader degenerate contribution.
Similar features are obtained for the polycrystalline case (bottom).
The TDDFT calculation displays 5 main features at 8, 16, 27, 38 and 46 eV, at positions close to the peaks A, B, E, I and FG by less than 2 eV (see also Fig.~\ref{figure-biblio}).
TDDFT reveals that the shoulder A is due to single particle transitions (O 2$p \to$ Hf 5$d$), like also F and G (Hf 5$p \to$ 5$d$); peak B is the only real bulk plasmon, while E and I are collective excitations.
Local-field effects severely damp peak I, which would otherwise be the main total plasmon.
Details about the theoretical calculation and interpretation will be provided elsewhere\cite{Hungetal.}.
Our TDDFT calculations with local-field effects can be considered the best-available simulated energy-loss spectra, with net improvement in agreement with experiment compared to previous DFT calculations\cite{ParkYang} (which did not predict the damping of the I collective excitation).

We now analyze differences in spectra measured along different crystal directions.
The most evident change is a modulation of the oscillator strengths of the shoulder A and in particular of the plasmon B.
To correctly interpret this finding, it is important to deconvolute the surface loss functions (right), since the most intense feature of the surface losses is located around 13 eV, close to the position of the bulk plasmon. Strong differences in surface losses may therefore contribute to the lack of perfect congruence among literature data.

When all parameters are well controlled, the comparison between experiment and the TDDFT is good. Although less pronounced than in the experiment, TDDFT confirms the plasmon modulation depending on the direction.
In both TDDFT and experiment, this is the most evident effect of anisotropy, the other changes being less obvious.

\begin{figure*}
\includegraphics[width=0.9\textwidth]{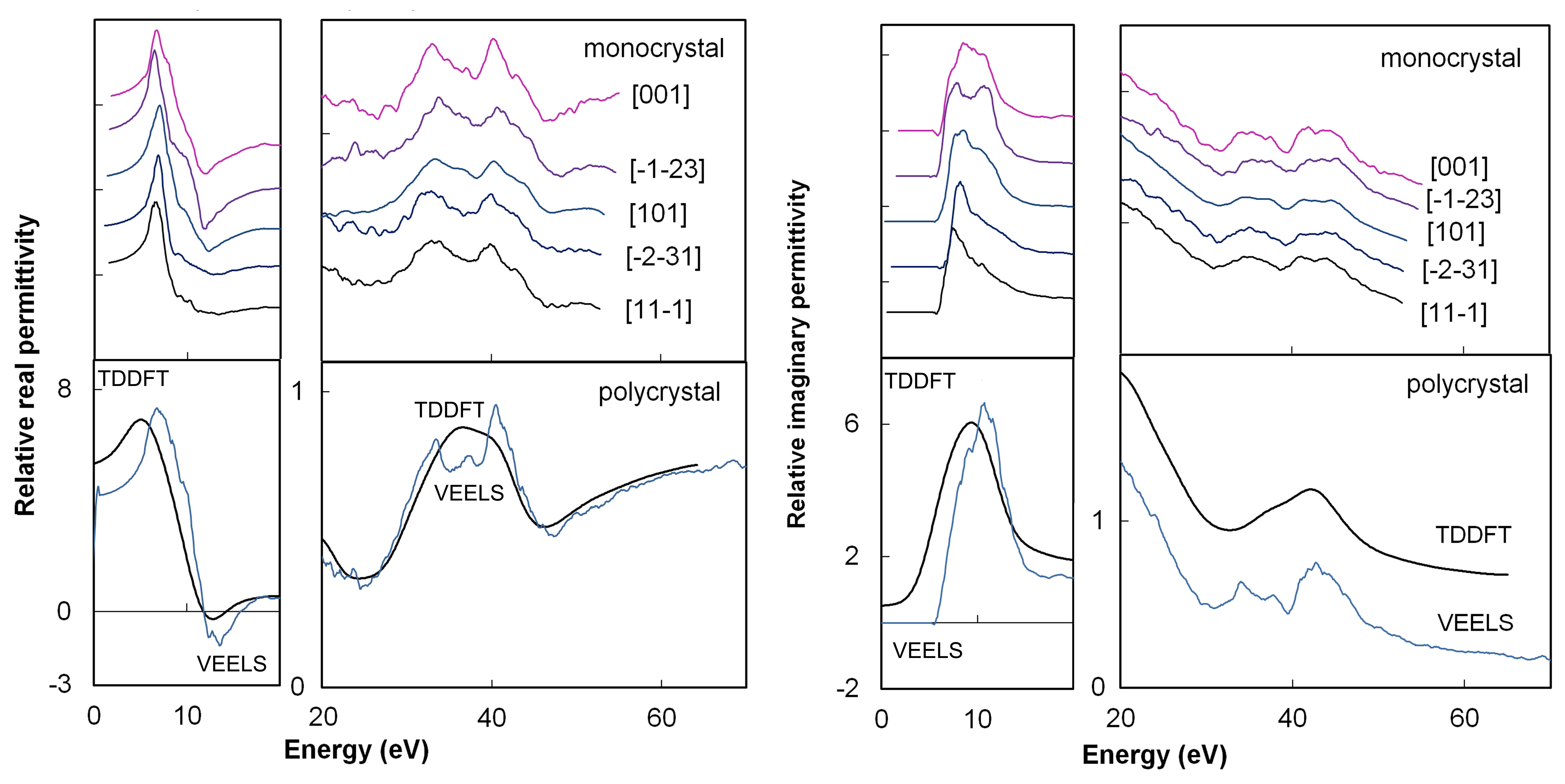}
\caption{
Real (left) and imaginary (right) permittivity of m-HfO$_2$ measured in the case of monocrystals (top) and polycrystal (bottom) at several momentum-transfer directions, in comparison with \textit{ab initio} TDDFT simulations.
Curves in the top panels are vertically shifted to facilitate comparison.
}
\label{anisotropy-epsilon-figure}
\end{figure*}

This is also evident in the TDDFT spectra shown in Fig.~\ref{figure-biblio} (top curves), where we plot the energy-loss along the three main crystallographic directions, as well as a directional averaged spectrum.
The most important anisotropy effect is observed once again on the bulk plasmon at 16 eV, with minor modifications on the rest of the spectrum.
The plasmon appears as a well separated peak in the [001] direction, while it reduces to a shoulder in the other directions.

To the best of our knowledge, this anisotropy in the dielectric properties and spectroscopy of m-HfO$_2$ has not been previously reported.
This can shed new light on the interpretation of VEELS spectra previously measured.
In Fig.~\ref{figure-biblio} we report the measured VEELS spectra of m-HfO$_2$ published in Refs.~[\onlinecite{CheynetPokrant,Agustin,Couillard,ParkYang}].
We also show our measured VEELS (polycrystalline sample).

It is evident that the measured VEELS present some disagreements among them.
We observe in particular a different attribution of the intensity of the plasmon at $\sim$16 eV, which is the most intense peak \textit{e.g.} in Cheynet \textit{et al.}\cite{CheynetPokrant,EELSAtlas}, while it is less intense than the collective excitation at $\sim$27 eV in Park and Yang\cite{ParkYang}.
We observe different intensities also on the 47 eV LF-damped plasmon.
In light of our findings, the effect of anisotropy could partially contribute to explain the differences among previously measured VEELS spectra, even if many additional experimental factors, like sample thickness, roughness, defectivity, could also affect the result. Performing absolute and perfectly quantitative measurements is indeed a real challenge. Current stability, convergence and collection angle, energy resolution, type of measurement (EFTEM vs. STEM), and the zero-loss removal method might have an impact on the resulting spectra, even if perfect single scattering configurations and highly accurate corrections of anisochromaticity and spatial drifts are used. The data analysis could also impact the Kramers-Kronig process significantly. 
Nevertheless, when limiting the influence of these factors, anisotropies in m-HfO$_2$ can be clearly observed, in agreement with TDDFT calculations.

The real and imaginary parts of the dielectric permittivity are represented in Fig.~\ref{anisotropy-epsilon-figure}.
Again, the overall agreement between experiment and simulation is good, in spite of the complexity of m-HfO$_2$.
In particular, the real permittivity passes zero near 15 eV, indicating that peak B is a proper bulk plasmon, in agreement with TDDFT.
Overall, we find good agreement between experiment and TDDFT-RPA (with local-field effects) even in the 20-70 eV range, which improves on previous DFT calculations\cite{ParkYang}.
It is important to note that some residual discrepancies are still present, especially in the intensity of the first plasmon peak, which is particularly sensitive to transitions possibly involving defect levels.
But the theoretical description can be already considered satisfactory in order to interpret and describe the correct physics in m-HfO$_2$.

Finally, we report our TDDFT calculated dielectric constants for m-HfO$_2$.
Here again we find significant anisotropies.
TDDFT predicts a dielectric constant of 5.0 in the [100] and [010] directions, but only 4.6 in the [001] $z$ direction, a difference of around 10\%.
The same 10\% difference is confirmed also using the TDDFT adiabatic LDA approximation\cite{LDA}.
This anisotropy in the dielectric properties, both in the static regime and also at the bulk plasmon frequency, might have possible technological applications.

\paragraph{Conclusion}
We have measured VEELS spectra of m-HfO$_2$ in correspondence to well defined momentum-transfer crystallographic directions.
We have compared the experimental spectra to TDDFT calculated, finding a good agreement.
This has allowed us to correctly interpret and understand the dielectric properties of polycrystalline or monocrystalline m-HfO$_2$.
We have found a significant anisotropy in the dielectric properties, mostly on the bulk plasmon at $\sim$16 eV and on the dielectric constant.
This can impact the behaviour of (opto)electronic devices based on this important high-$\kappa$ material.

\paragraph{Acknowledgements}

We acknowledge support from ETSF for simulations and the nanocharacterisation platform (PFNC, \url{http://www.minatec.org/pfnc-plateforme-nanocaracterisation}) for experiments, and Cezus (Areva) for providing m-HfO$_2$ powder.
Computer time has been provided by the French GENCI supercomputing center (projects i2012096-655 and 544).


\begin{thebibliography}{}
\bibitem{conv-param}
The DFT calculation is converged at a kinetic energy cut-off of 150 Ha.
In the TDDFT calculation we used 300 bands, 9475 plane waves to represent wavefunctions and 329 for polarizability and dielectric matrices.
We sampled the Brillouin zone by a 4$\times$4$\times$4 shifted k-points grid.
\end{thebibliography}
\end{document}